\documentclass[a4paper,fleqn,usenatbib]{mnras}

\usepackage[T1]{fontenc}
\usepackage{ae,aecompl}
\usepackage[utf8]{inputenc}

\usepackage{graphicx}	
\usepackage{amsmath}	
\usepackage{amssymb}	
\usepackage{import}
\usepackage{lmodern}
\RequirePackage{fix-cm}

\usepackage{numprint}
\usepackage{SIunits}

\usepackage{grffile}

\newcommand{\er}{\mathbf{e}_{\rm r}}
\newcommand{\ex}{\mathbf{e}_{\rm x}}
\newcommand{\ey}{\mathbf{e}_{\rm y}}
\newcommand{\ez}{\mathbf{e}_{\rm z}}

\newcommand{\rlight}{r_{\rm L}}

\title[FFE off-centred dipoles]{Off-centred force-free neutron star magnetospheres}

\author[J. P\'etri]{
J. P\'etri\thanks{E-mail: jerome.petri@astro.unistra.fr}
\\
Universit\'e de Strasbourg, CNRS, Observatoire astronomique de Strasbourg, UMR 7550, F-67000 Strasbourg, France.
}

\date{Accepted XXX. Received YYY; in original form ZZZ}

\pubyear{2020}

\begin{document}
\label{firstpage}
\pagerange{\pageref{firstpage}--\pageref{lastpage}}
\maketitle

\begin{abstract}
Neutron star electromagnetic activity produces pairs that fill their magnetosphere represented to the zeroth order by the force-free approximation. Neither dissipation nor acceleration nor radiation from charged particles is expected from this simplified model. So far, only centred dipole magnetic fields have been studied in this limit. In this paper, we explore the consequences of a rotating off-centred dipole on the force-free magnetosphere, showing the new magnetic field geometry, its spin-down luminosity as well as the electromagnetic kick and torque felt by the neutron star. Solutions are obtained by time-dependent numerical simulations of the force-free regime using our pseudo-spectral code written in spherical coordinates. Our results are also compared to known analytical expressions found for the off-centred vacuum dipole by an expansion to lowest order in the parameter $\epsilon = d/R$, where $d$ is the displacement of the dipole from the stellar centre and $R$ the neutron star radius. The presence of a force-free plasma enhances the spin-down luminosity as well as the electromagnetic kick and torque with respect to a centred force-free dipole. The impact on isolated and binary neutron stars is revised in light of these new results.
\end{abstract}

\begin{keywords}
	magnetic fields - methods: numerical - stars: neutron - stars: rotation - pulsars: general - plasmas.
\end{keywords}



\section{Introduction}

Neutron star magnetic fields are usually idealized by a centred dipole rotating in vacuum. However, because of the ultra-strong rotating magnetic field, a huge electric field is induced, that expels charged particles from the surface, filling the magnetosphere with electron/positron pairs. The simplest description is a force-free magnetosphere where the plasma exactly screens the electric field component parallel to the magnetic field. Detailed numerical investigations of this force-free magnetosphere have been undertaken numerically, first for an axisymmetric rotator using an iterative scheme to solve the Grad-Shafranov equation \citep{contopoulos_axisymmetric_1999}, then for the full 3D oblique rotator by using time-dependent simulations \citep{spitkovsky_time-dependent_2006, petri_pulsar_2012}. Even particle in cell simulations were performed to account for possible gaps and acceleration of particles \citep{cerutti_particle_2015, kalapotharakos_three-dimensional_2018}. However, in all these works, the magnetic dipole was assumed to lie at the stellar centre. There is however no particular reason to start with such an hypothesis as the internal magnetic field is certainly neither truly dipolar nor exactly centred. Electric currents within the star could easily break any symmetry, producing fields that are non dipolar and off-centred. For neutron stars shining as pulsars in radio and thermal X-rays, like PSR~J1136+1551, this asymmetry is indirectly observed as a time lag between the radio pulse profile and the X-ray thermal peak emission \citep{petri_joint_2020} allowing to constrain the geometry of the off-centred dipole. The latter also impacts the dynamics of the neutron star, producing a high velocity kick as shown by \cite{harrison_acceleration_1975} and by \cite{tademaru_acceleration_1976} for a dipole evolving in vacuum.

The recent discovery by NICER of a complex magnetic structure at the surface of the millisecond pulsar PSR~J0030+0451 shows the necessity to add multipolar components and in particular a significant quadrupole part \citep{bilous_nicer_2019}. This magnetic field is filled with electron-positron pairs, producing a feedback current that is approximated in the ideal plasma case by a force-free expression (FFE). Adding small scale magnetic field structures in the neutron star magnetosphere becomes compulsory to interpret the increasing amount of accurate observations in radio, X-ray and gamma-ray. The off-centred dipole offers a simple picture to naturally include such higher order terms.

The vacuum rotating off-centred dipole has been carefully investigated by \cite{petri_impact_2019}, showing the good agreement between analytical results and numerical simulations. He also discussed possible implications for binary neutron star systems, especially for the eccentricity of their orbit. However, a vacuum magnetosphere is unrealistic because of the presence of pairs surrounding the dipole. Thus these conclusions must be reinvestigated in light of a plasma filled magnetosphere starting with the force-free approximation as we show below.

In this paper we compute numerical solutions for the electromagnetic field in the force-free regime outside an off-centred rotating dipole. The off-centred rotating dipole model is exposed in Sec.~\ref{sec:Modele}. Sec.~\ref{sec:LigneChamp} shows two examples of magnetic field line geometries for particular orientations of the dipole. A complete set of simulations is summarised in Sec.~\ref{sec:Luminosite} for the spin-down luminosity and compared with its vacuum analogue. Such comparisons are extended  to the electromagnetic kick in Sec.~\ref{sec:Kick} and to the electromagnetic torque in Sec.~\ref{sec:Couple}. A short discussion about the revised impact on binaries containing neutron stars is highlighted in Sec.~\ref{sec:Binaire}. Conclusions are drawn in Sec.~\ref{sec:Conclusion}.

\section{The model}
\label{sec:Modele}

\subsection{Off-centred magnetic dipole}

The off-centred magnetic dipole has been introduced by \cite{petri_radiation_2016} and used in \cite{petri_impact_2019} to study the secular evolution of neutron stars. For completeness, we briefly remind the geometrical set up and notations used in these works.

The off-centred dipole is anchored in a perfectly conducting sphere of radius~$R$ in solid body rotation at an angular rate~$\Omega$. Its magnetic moment~$\bmu$ is located inside the star at a position given by the vector
\begin{equation}
\mathbf{d} = d \, (\sin \delta \, \cos(\Omega\,t) \, \ex + \sin \delta \, \sin (\Omega\,t) \, \ey + \cos \delta \, \ez )
\end{equation}
$d=\|\mathbf{d}\|$ being the distance from the stellar centre, $\delta$ the colatitude and $(\ex,\ey,\ez)$ a Cartesian orthonormal basis. The displacement is normalised by introducing the parameter $\epsilon=d/R<1$. The magnetic moment~$\bmu$ is directed along a unit vector~$\mathbf{m}$ identified by the angles~$(\alpha,\beta)$ such that
\begin{equation}
\mathbf{m} = \sin \alpha \, \cos (\beta+\Omega\,t) \, \ex + \sin \alpha \, \sin (\beta+\Omega\,t) \, \ey + \cos \alpha \, \ez .
\end{equation}
Inside the star, the magnetic field is assumed to be a static dipole shifted by $\mathbf{d}$ \citep{burnett_stokes_2014}
\begin{equation}
\label{eq:OffcenteredDipole}
\mathbf{B} = \frac{B\,R^3}{\|\mathbf{r} - \mathbf{d}\|^3} \, \left[ \frac{3\,\bmath{\mu} \cdot (\mathbf{r} - \mathbf{d})}{\|\mathbf{r} - \mathbf{d}\|^2} \, (\mathbf{r} - \mathbf{d}) - \bmath{\mu} \right]
\end{equation}
where $B$ is the surface magnetic field strength at the magnetic equator and $\mathbf{r}$ the position vector.
We emphasize that the field inside the star is certainly not given by the displaced point dipole as prescribed in Eq.~\eqref{eq:OffcenteredDipole}. However, this might be a reasonable approximation to the field on the neutron star surface that we use as a boundary condition to compute the force-free field outside the star.

Taking the surface boundary conditions from the static solution Eq.~\eqref{eq:OffcenteredDipole}, we solved the time dependent Maxwell equations in the force-free regime by using our pseudo-spectral code detailed in \cite{petri_general-relativistic_2014}. These boundary conditions are given by the continuity of the radial component of the magnetic field~$B_{\rm r}$ and the tangential component of the electric field. In the corotating frame inside the star, this leads to a vanishing electric field $\mathbf{E}' = \mathbf{0}$ where unprimed coordinates and fields are evaluated in the inertial frame, and primed ones are evaluated in the rotating frame. For the outer boundary, we impose outgoing waves. Compared to previous simulations, we now solve Maxwell equations in a corotating coordinate system as explained in the next paragraph. This ensures a stationary state to which the solution has to relax.

\subsection{Maxwell equations in a rotating coordinate system}

We are looking for a stationary solution of the electromagnetic field which is actually static in the frame corotating with the star. An observer can only corotate with the star up to the light-cylinder. Measuring the electromagnetic field is impossible for such an observer outside the light-cylinder because the metric has no physical significance any more. It is impossible to describe the electrodynamics in whole space with the field locally measured by a corotating observer because it does not exist when $r \geq \rlight$. A rotating frame is also not easily defined \citep{koks_simultaneity_2019} contrary to a rotating coordinate system that does not require to move slower than the speed of light without contradicting special relativity.

We found it however useful to keep the definition of the electromagnetic field as measured in the inertial reference frame but using a rotating cylindrical coordinate system~$(t'=t,r'=r,\phi'=\phi-\Omega\,t,z'=z)$ (remember that unprimed quantities are given in the inertial frame and primed quantities in the rotating frame.) In such a case the time derivative of any vector field~$\mathbf{A}$ is given by
\begin{equation}
\label{eq:Derivee}
\frac{\partial \mathbf A}{\partial t} = \frac{\partial \mathbf A}{\partial t'} + \textrm{curl} \, ( \mathbf{V}_{\rm rot} \wedge \mathbf A ) - \mathbf{V}_{\rm rot} \, \textrm{div} \mathbf A .
\end{equation}
The solid body corotation velocity, expressed in the inertial frame, is simply
\begin{equation}
\label{eq:VitesseCorotation}
\mathbf{V}_{\rm rot} = \boldsymbol{\Omega} \wedge \mathbf{r} = r\,\Omega\,\mathbf{e}_\phi .
\end{equation}
With the correspondence established in eq.~(\ref{eq:Derivee}), in the rotating coordinate system, Maxwell equations become
\begin{subequations}
	\label{eq:MaxwellCorotation}
	\begin{align}
	\frac{\partial \mathbf{B}}{\partial t'} & = - \, \textrm{curl} \, (  \mathbf{E} + \mathbf{V}_{\rm rot} \wedge \mathbf B ) \\
	\frac{\partial \mathbf{E}}{\partial t'} & = \textrm{curl} \, ( c^2 \, \mathbf{B} - \mathbf{V}_{\rm rot} \wedge \mathbf E ) - \frac{\mathbf{j}}{\varepsilon_0} + \mathbf{V}_{\rm rot} \, \textrm{div} \mathbf E .
	\end{align}
\end{subequations}
The force-free current density is given solely by the electromagnetic field according to \citep{blandford_lighthouse_2002}
\begin{equation}
\label{eq:J_Ideal}
\mathbf j = \rho_{\rm e} \, \frac{\mathbf{E}\wedge \mathbf{B}}{B^2} + \frac{\mathbf{B} \cdot \nabla \wedge \mathbf{B} / \mu_0 - \varepsilon_0 \, \mathbf{E} \cdot \nabla \wedge \mathbf{E}}{B^2} \, \mathbf{B} 
\end{equation}
the electric charge density being
\begin{equation}\label{eq:MaxwellGauss}
\varepsilon_0 \, \nabla \cdot \mathbf{E} = \rho_{\rm e} .
\end{equation}
Note however the subtleties that $\mathbf{E}$ and $\mathbf{B}$ are still defined as observed in the inertial frame therefore they remain unprimed quantities. There is no particular problem at the light-cylinder when Maxwell equations are written in this way. In the next sections, we use this formulation to solve for the force-free magnetosphere for an oblique rotator.

In order to emphasize the role of each angle $\alpha$, $\beta$, $\delta$ and the displacement~$d$ on the electromagnetic field, a full set of runs have been performed. Results of the simulations are synthesised by plotting the magnetic field, the spin-down luminosity and the associated electromagnetic kick and torque.

\begin{figure*}
	\centering
	\begin{tabular}{cc}
		\includegraphics[width=0.5\linewidth]{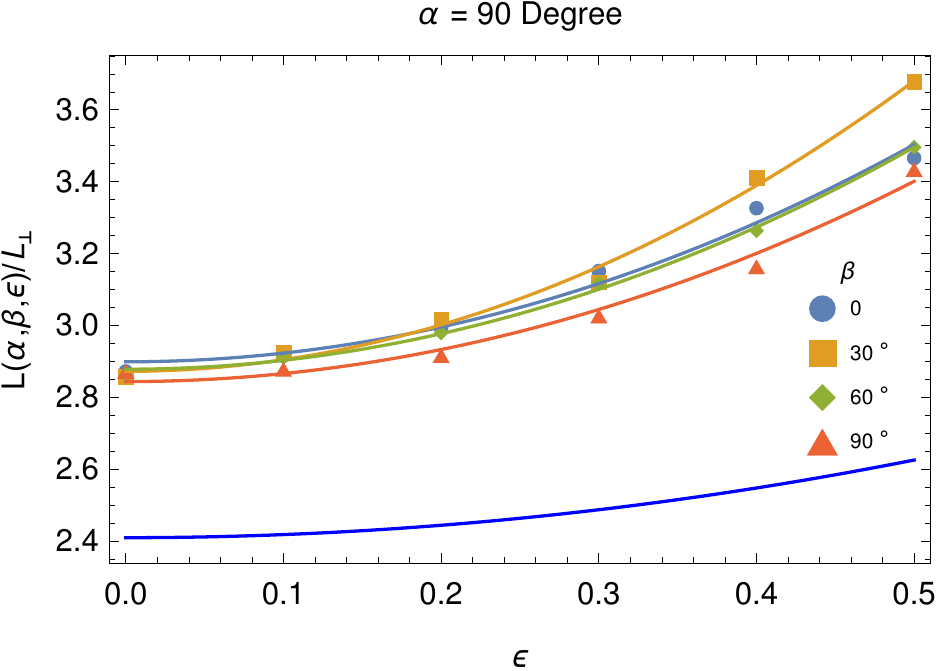} &
		\includegraphics[width=0.5\linewidth]{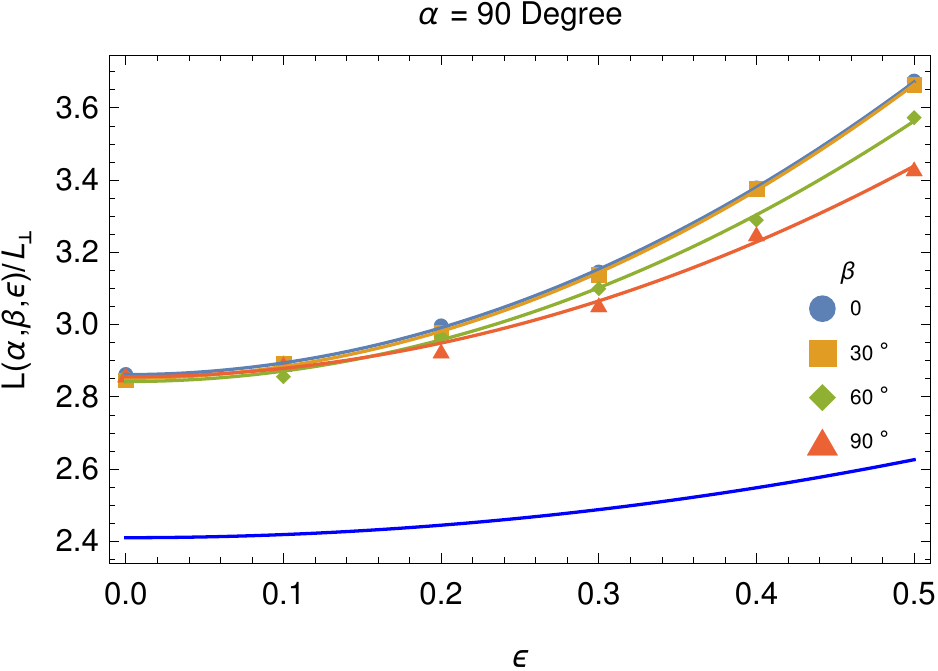} \\	
	\end{tabular}
	\caption{Spin-down luminosity depending on the grid resolution for $\alpha=90\degr$, $\delta=90\degr$ and varying~$\beta$ and $\epsilon$. The resolution is $N_r \times N_\theta \times N_\phi= 65 \times 16 \times 32$ on the left and $N_r \times N_\theta \times N_\phi= 129 \times 32 \times 64$ on the right. The blue line corresponds to the function $(3/2+L_{\rm vac}/L_\perp)$ and the colour solid lines to a fit of the FFE simulations.}
	\label{fig:luminosite_resolution}
\end{figure*}

In order to speed up the computation of the large 4D space parameter, we used an artificially high spin rate given by $a=R/\rlight=0.3$ for the whole set of simulations. Moreover, we only computed solutions for $\delta=90\degree$ in all the results shown below. Going to slower rotation rates will not qualitatively change our main results. The spatial resolution in the spherical grid $(r\,\theta,\phi)$ is given by $N_r \times N_\theta \times N_\phi= 129 \times 32 \times 64$. We checked that this grid is sufficient to achieve a good accuracy by computing solutions with a lower grid resolution of $N_r \times N_\theta \times N_\phi= 65 \times 16 \times 32$. The impact of this coarser resolution on the spin-down luminosity is shown on left plot of Fig.~\ref{fig:luminosite_resolution} for $\alpha=90\degr$ and varying~$\beta$ and $\epsilon$ and compared to the finer grid on the right plot of Fig.~\ref{fig:luminosite_resolution}. $L_{\rm vac}$ represents the vacuum off-centred dipole spin-down as given in \cite{petri_radiation_2016} and $L_\perp$ the spin-down for an orthogonal point dipole rotating in vacuum, see Eq.\eqref{eq:L_perp}. The discrepancies are indeed not appreciable.

\section{Field lines}
\label{sec:LigneChamp}

Plotting the magnetic field lines gives a first insight into the impact of a rotating off-centred force-free dipole. It is not possible to show all geometrical configurations, so we focus on two particular geometries where some field lines are entirely contained in the equatorial plane. A perpendicular rotator with $\alpha = \delta = 90\degr$ is a good choice. We allowed some freedom in the displacement~$d$ and angle~$\beta$. As expected, for small off-centring~$d\ll R$ the field line structure resembles to the centred force-free dipole. The two armed spiral, reminiscent of a $\ell=1$ mode, is clearly visible, dragged by the stellar rotation at a constant speed~$\Omega$.

As a concrete example we choose $\beta=0\degr$ and $\epsilon=0.3$, obtaining the field lines shown in Fig.~\ref{fig:LigneChampB0} in red solid line and compared to the centred force-free dipole in blue dashed line~$\epsilon=0$. The off-centred configuration introduces some multipolar components of order~$\ell>1$ that decrease with radius~$r$ faster than the dipole. Therefore an observer located at large distances from the star $r\gg\rlight$, cannot notice the difference between centred and off-centred dipole because all higher multipoles become negligible. As a corollary, it is impossible to deduce the geometry of the dipole simply by reporting the field at large distances. Asymmetries only impact the stellar close neighbourhood.
\begin{figure}
	\centering
	\caption{Magnetic field lines of an off-centred force-free dipole with $\alpha=90\degr$, $\beta=0\degr$, $\delta=90\degr$ and $\epsilon=0.3$ (red solid line) compared to the centred solution $\epsilon=0$ (blue dashed line).}
	\label{fig:LigneChampB0}
\end{figure}

\begin{figure}
	\centering
	\includegraphics{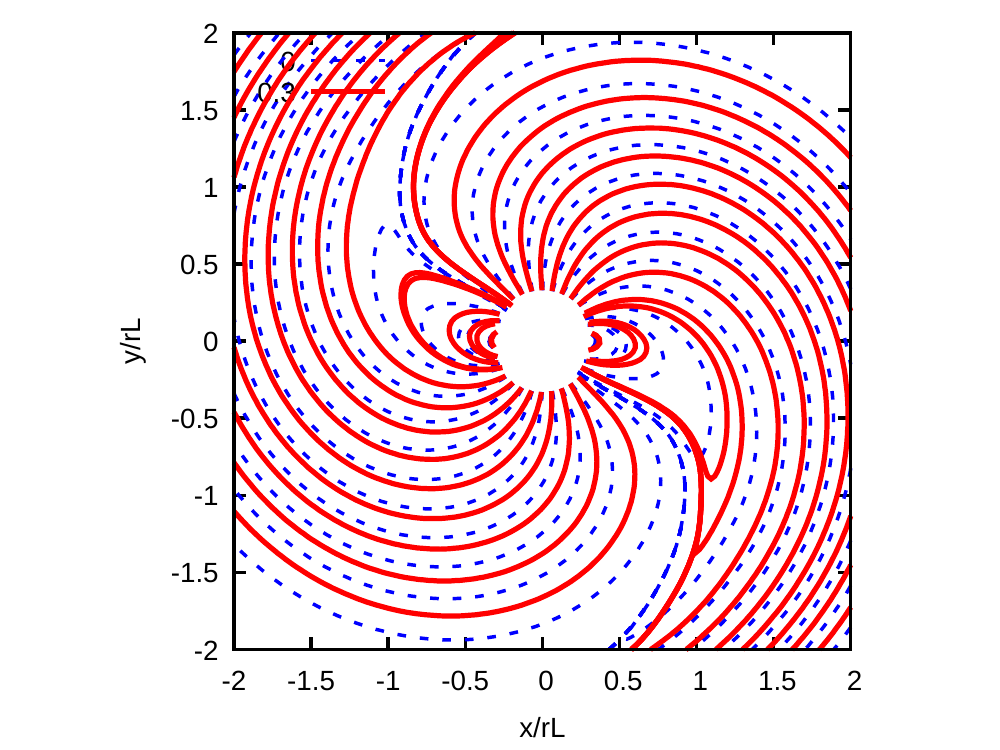}
	\caption{Magnetic field lines for an off-centred force-free dipole with $\alpha=90\degr$, $\beta=90\degr$, $\delta=90\degr$ and $\epsilon=0.3$ (red solid line) compared to centred solution $\epsilon=0$ (blue dashed line).}
	\label{fig:LigneChampB90}
\end{figure}

A second and similar example is shown in Fig.~\ref{fig:LigneChampB90} for $\beta=90\degr$ and $\epsilon=0.3$ in red solid line and can be compared to the centred force-free dipole in blue dashed line. The same conclusions as before apply except that now, outside the light-cylinder, a shift in phase appears for the spiral structure, with respect to the centred case. However, this shift remains too weak to be measured. The only way to deduce the true magnetic geometry requires indirect measurement, investigating its electromagnetic activity and emission properties close to the surface.
We continue our study by computing more quantitative physical parameters like the spin-down luminosity, the electromagnetic kick and its associated torque.

\section{Spin-down luminosities}
\label{sec:Luminosite}

Isolated neutron stars slow down due to electromagnetic radiation. It is quantified by the spin-down luminosity, braking the neutron star rotation. Since the work of \cite{deutsch_electromagnetic_1955} an exact expression is known for a dipole in vacuum. It has recently been extended by \cite{petri_radiation_2016} for an off-centred dipole, giving approximate formulas for the dipole and quadrupole contributions that are
\begin{subequations}
	\label{eq:Lm1e2}
	\begin{align}
	L_{m=1} & = L_\perp \, \left[ \left( 1 - a^2 \right) \, \sin^2\alpha + \frac{24}{25} \, a^2 \, \epsilon^2 \, \cos^2\alpha \right] \\
	L_{m=2} & = \frac{48}{5} \, L_\perp \, a^2\,\epsilon^2 \, \sin^2\alpha \ .
	\end{align}
\end{subequations}
where $a=R/\rlight$ and the centred perpendicular point dipole spin-down in vacuum is
\begin{equation}
\label{eq:L_perp}
L_\perp = \frac{8\,\upi}{3\,\mu_0\,c^3} \, \Omega^4 \, B^2 \, R^4 .
\end{equation}
Expressions~(\ref{eq:Lm1e2}) assume that $\delta=90\degr$. The general case with arbitrary angle~$\delta$ can be found but is too lengthy to show and in the present work we only consider $\delta=90\degr$. These expression have been confirm by direct time-dependent numerical simulations performed by \cite{petri_impact_2019}.

Now the presence of the plasma changes this formal geometrical dependence. Our new results depend on the angle~$\beta$ contrary to what is expected from eq.~(\ref{eq:Lm1e2}) for vacuum. We performed a set of runs with relevant geometric parameters by varying the set $(\alpha, \beta, \epsilon)$ and choosing different rotation periods symbolized by the adimensionalized parameter~$a$.

From the simulations we calculate the spin-down luminosity~$L_{\rm FFE}$ by integrating the radial component of the Poynting flux $\mathbf{S} = \mathbf{E} \times \mathbf{B}/\mu_0$ on a sphere of radius equal to the light-cylinder radius~$\rlight$
\begin{equation}\label{eq:L_FFE}
L_{\rm FFE} = \oint_{\mathcal{S}_{\rm L}} (\mathbf{S} \cdot \er) \, r^2 \,  d\Omega_{\rm L} = \rlight^2 \, \int_0^\upi \int_0^{2\upi} S_{\rm r} \, \sin\theta \, d\theta \, d\phi
\end{equation}
where $\mathcal{S}_{\rm L}$ is the sphere of radius~$\rlight$ and $d\Omega_{\rm L}$ the solid angle subtended by this sphere and expressed in spherical polar coordinates $(\theta,\phi)$ as $d\Omega_{\rm L}=\sin\theta \, d\theta \, d\phi$. Ideally integration on any sphere of arbitrary radius~$r$ should give the same results because of electromagnetic energy conservation in the force-free regime, but due to numerical dissipation, outside the light-cylinder, exact energy conservation is violated because of the presence of a current sheet that acts as a sink of energy. The electromagnetic force is computed in a similar way by integrating along the same sphere the component of the kick along the rotation axis, i.e. along $\ez$
\begin{subequations}\label{eq:F_z}
\begin{align}
F_{\rm z} & = \frac{1}{c} \oint_{\mathcal{S}_{\rm L}} (\mathbf{S} \cdot \ez) \, r^2 \,  \cos \theta \, d\Omega_{\rm L} \\
 &= \frac{\rlight^2}{c} \, \int_0^\upi \int_0^{2\upi} S_{\rm z} \, \cos \theta \, \sin\theta \, d\theta \, d\phi
\end{align}
\end{subequations}
see \cite{petri_radiation_2016}.

\begin{figure*}
	\centering
	\begin{tabular}{cc}
	\includegraphics[width=0.49\linewidth]{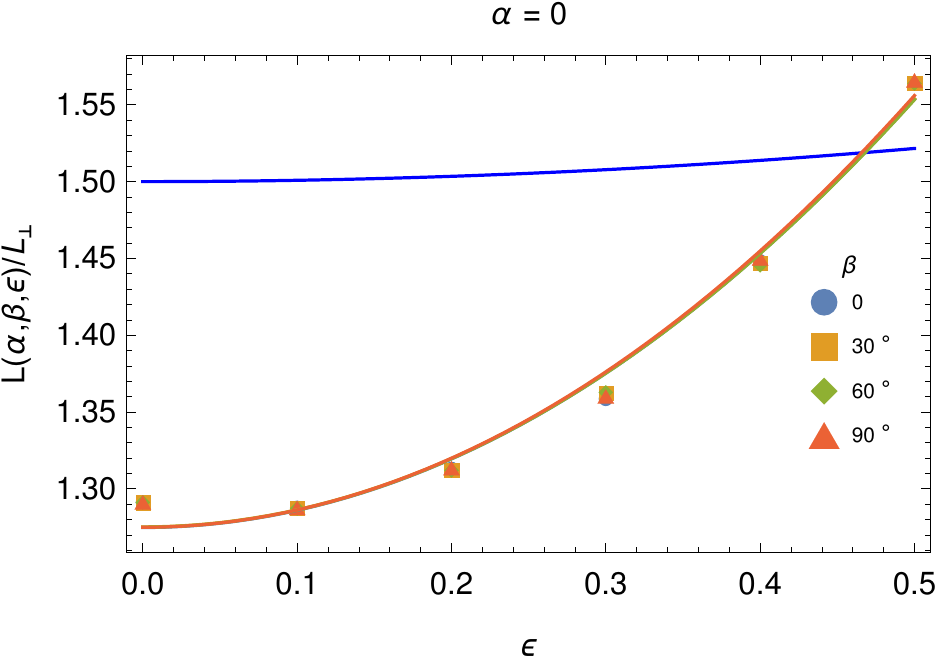} &
	\includegraphics[width=0.49\linewidth]{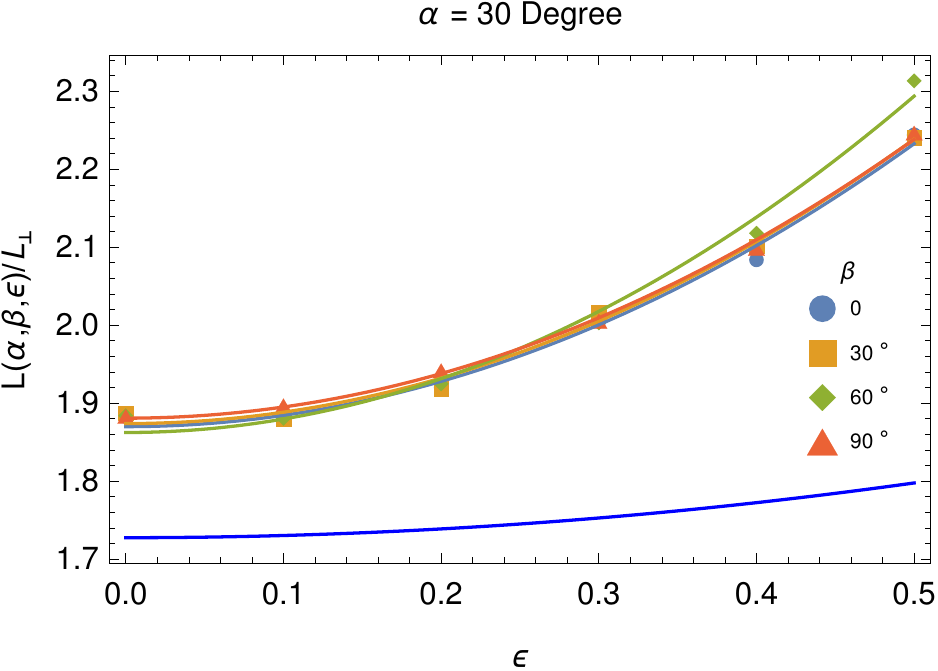} \\
	\includegraphics[width=0.49\linewidth]{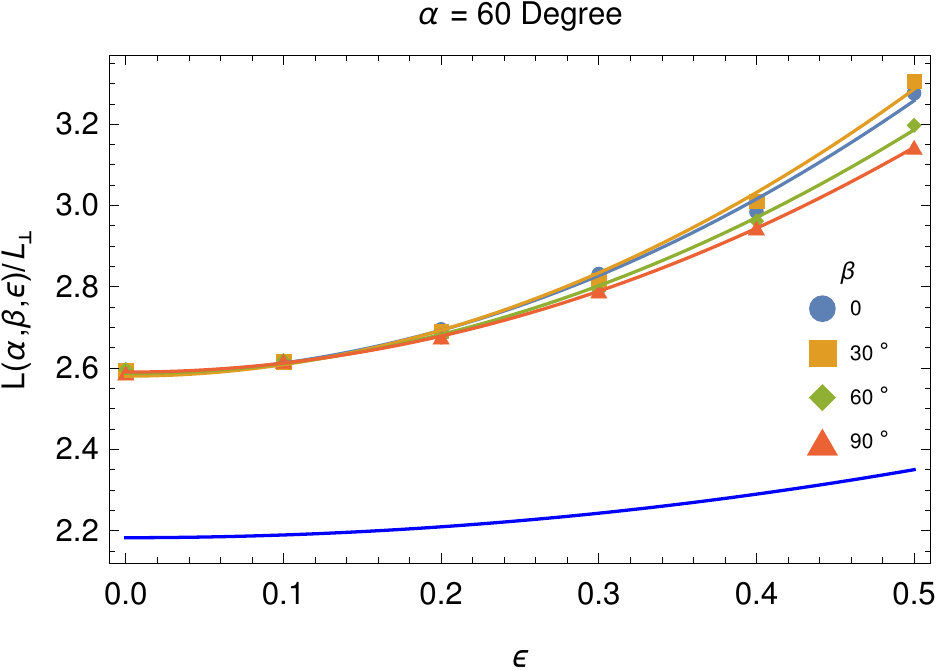} &
	\includegraphics[width=0.49\linewidth]{luminosite_r0.3_a90_n129_nt32_0.2_depl.pdf}
	\end{tabular}
	\caption{Variation of the spin-down luminosity depending on the displacement $\epsilon$, obliquity $\alpha$ and $\beta$ for $a=0.3$, and marked as coloured symbols. The blue line corresponds to the function $(3/2+L_{\rm vac}{/L_\perp})$ and the colour solid line to a fit of the FFE simulations.}
	\label{fig:luminosite_depl}
\end{figure*}

\begin{figure}
	\centering
	\includegraphics[width=0.99\linewidth]{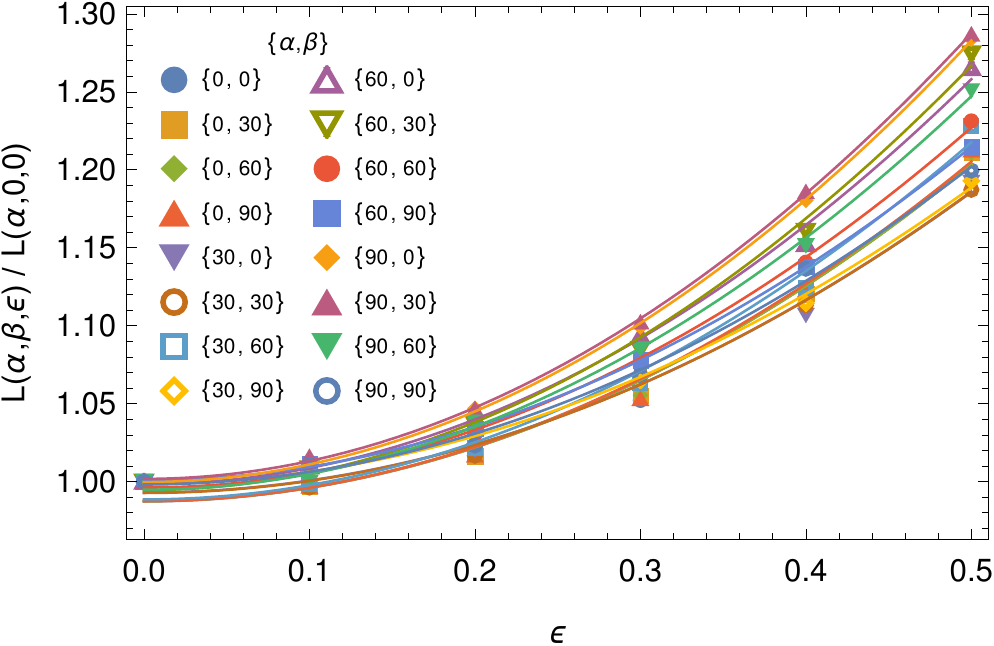}
	\caption{Normalized spin-down luminosity depending on the displacement $\epsilon$, obliquity $\alpha$ and $\beta$ for $a=0.3$, and marked as coloured symbols, see the legend for the labels corresponding to different couples~$\{\alpha,\beta\}$ in degrees. Solid coloured lines are best fits.}
	\label{fig:luminosite_depl_norm}
\end{figure}

For $a=0.3$ we summarize the simulation outputs for $\alpha=\{0\degr, 30\degr, 60\degr, 90\degr\}$ and $\beta=\{0\degr, 30\degr, 60\degr, 90\degr\}$ in Fig.~\ref{fig:luminosite_depl}. For comparison, we add the vacuum spin-down expectations in blue solid lines, offset by an amount 1.5 for ease of readability, and a fit to the FFE simulations in colour solid lines associated to $\beta$.
Fig.~\ref{fig:luminosite_depl_norm} shows the normalized spin-down luminosity evolution with respect to the displacement~$\epsilon$ for fixed $\alpha$, $\beta$ and $\delta$. We observe that with this normalisation the luminosity behaviour is rather insensitive to $\alpha$ and $\beta$ and well represented by a mean fit given by
\begin{equation}\label{eq:luminosite_fit}
L(\alpha, \beta, \epsilon) \approx ( 0.994 + 0.918 \, \epsilon^2 ) \, L(\alpha, 0, 0)
\end{equation}
where $L(\alpha, 0, 0)$ corresponds to the spin-down luminosity for the centred FFE dipole.
A large displacement pushing the magnetic moment close to the surface with $\epsilon \approx 1$ almost doubles the spin-down luminosity with respect to a centred dipole.

\section{Electromagnetic kick}
\label{sec:Kick}

\begin{figure}
	\centering
	\includegraphics[width=0.99\linewidth]{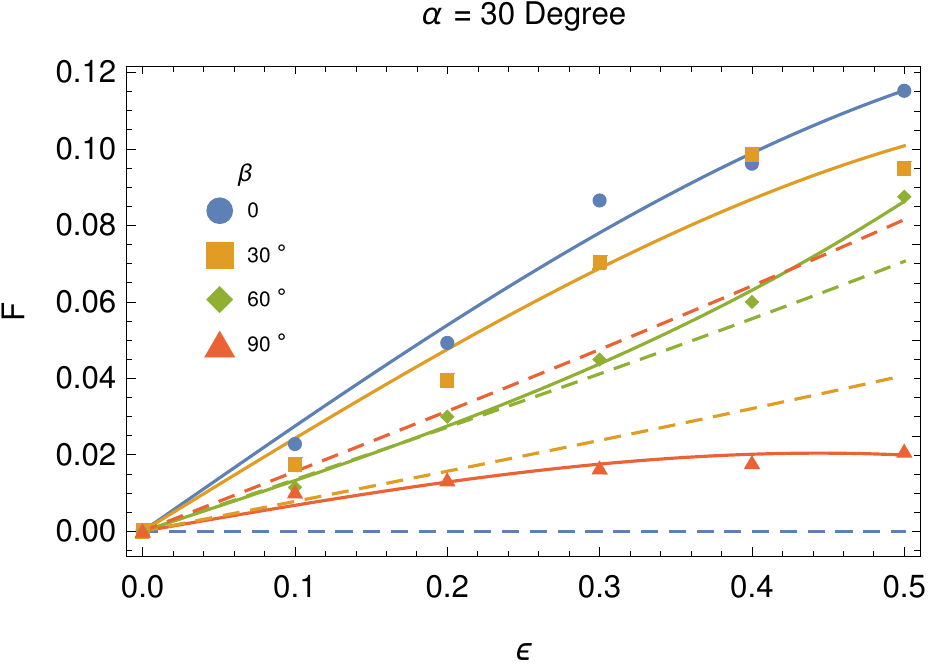}
	\caption{Electromagnetic force induced by a rotating off-centred dipole for different displacements~$\epsilon$ and different angles~$\beta$ for $\alpha=30\degr$, and marked as coloured symbols. Solid coloured lines are best fits. The dashed coloured lines correspond to the vacuum analogue.}
	\label{fig:force_a30_depl}
\end{figure}

\begin{figure}
	\centering
	\includegraphics[width=0.99\linewidth]{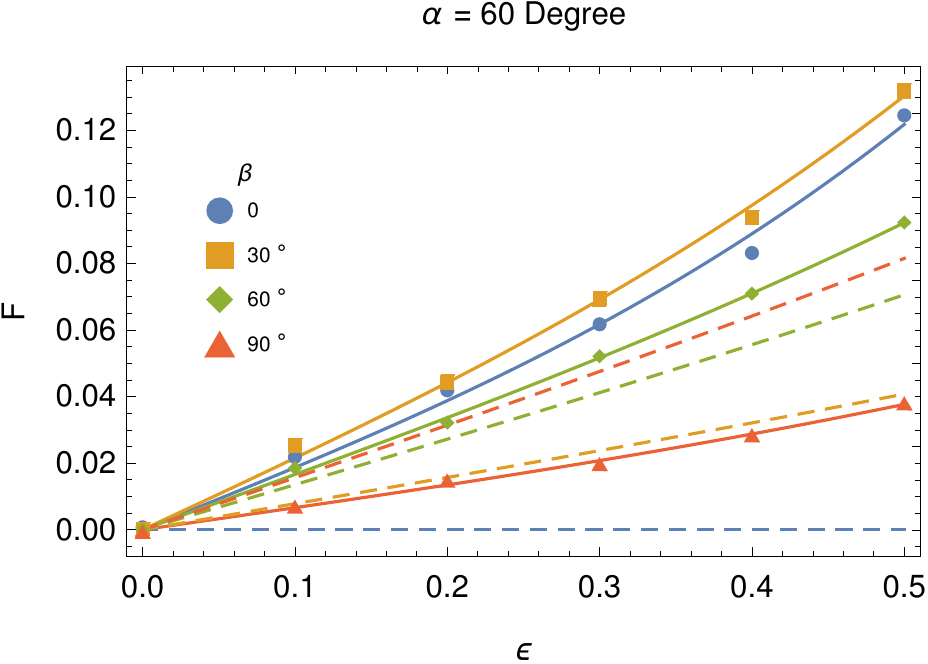}
	\caption{Electromagnetic force induced by a rotating off-centred dipole for different displacements~$\epsilon$ and different angles~$\beta$ for $\alpha=60\degr$, and marked as coloured symbols. Solid coloured lines are best fits. The dashed coloured lines correspond to the vacuum analogue.}
	\label{fig:force_a60_depl}
\end{figure}

As for the spin-down luminosity, it is interesting to compare force-free and vacuum electromagnetic kick results. Therefore we remind the kick expressions for the dipole~$m=1$ and quadrupole~$m=2$ contributions as found in \cite{petri_radiation_2016} and valid for $\delta=90\degr$. They read respectively
\begin{subequations}
	\begin{align}
	F_{m=1} & = \frac{6}{5} \, \frac{L_\perp}{c} \, a\,\epsilon \, \cos\alpha \, \sin\alpha \, \sin\beta \\
	F_{m=2} & = \frac{256}{105} \, \frac{L_\perp}{c} \, a^3\,\epsilon^3 \, \cos\alpha \, \sin\alpha \, \sin\beta  .
	\end{align}
\end{subequations}
Following the same lines as in \cite{petri_impact_2019}, the electromagnetic kick is deduced from our new set of runs and compiled in Fig.~\ref{fig:force_a30_depl} for $\alpha=30\degr$ and in Fig.~\ref{fig:force_a60_depl} for $\alpha=60\degr$. A linear scaling with respect to the displacement~$\epsilon$ is found to good accuracy. Nevertheless, because $a=0.3$ we also added a $\epsilon^3$ term in the fits shown as solid colour lines. We also compare these results to the vacuum electromagnetic force in dashed coloured lines.
Now compared to a vacuum dipole, although both plots remain similar, we notice that the $\alpha=60\degr$ leads to slightly smaller electromagnetic forces compared to  $\alpha=30\degr$. A substantial difference with the vacuum rotator is the presence of a significant force even for $\beta=0\degree$ as soon as the shifted dipole operates with $\epsilon\geq0.1$. The $\sin\beta$ dependence is lost and a kick is expected for any orientation of the dipole contrary to the vacuum case.

\begin{figure}
	\centering
	\includegraphics[width=0.99\linewidth]{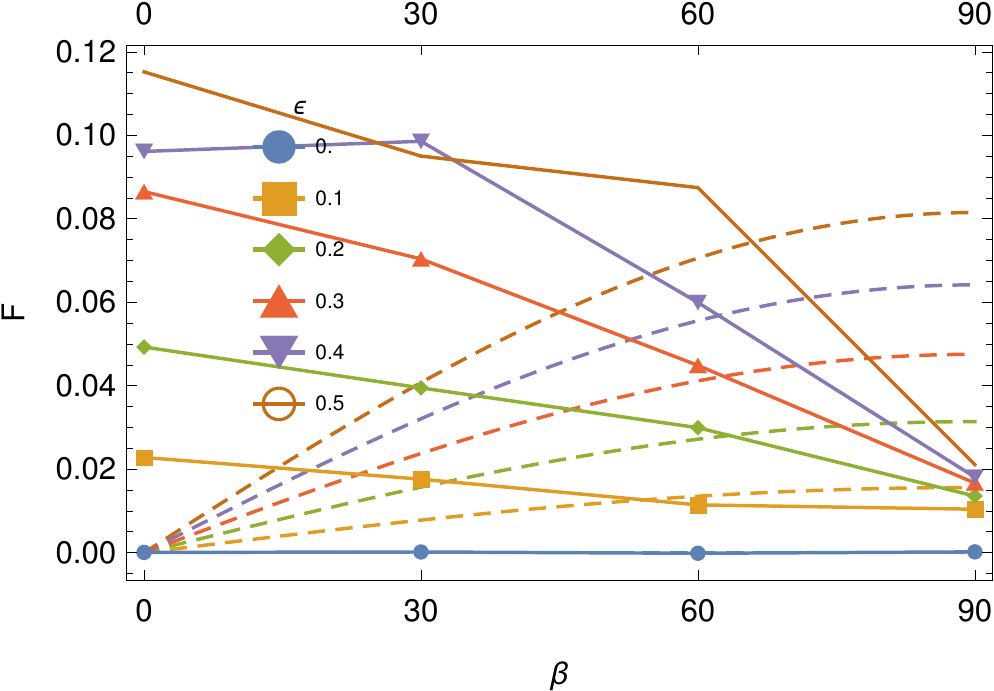}
	\caption{Electromagnetic force induced by a rotating off-centred dipole for different displacements~$\epsilon$ and different angles~$\beta$ for $\alpha=30\degr$, and marked as solid coloured lines. The dashed coloured lines correspond to the vacuum analogue. No fits are shown.}
	\label{fig:force_a30_beta}
\end{figure}

\begin{figure}
	\centering
	\includegraphics[width=0.99\linewidth]{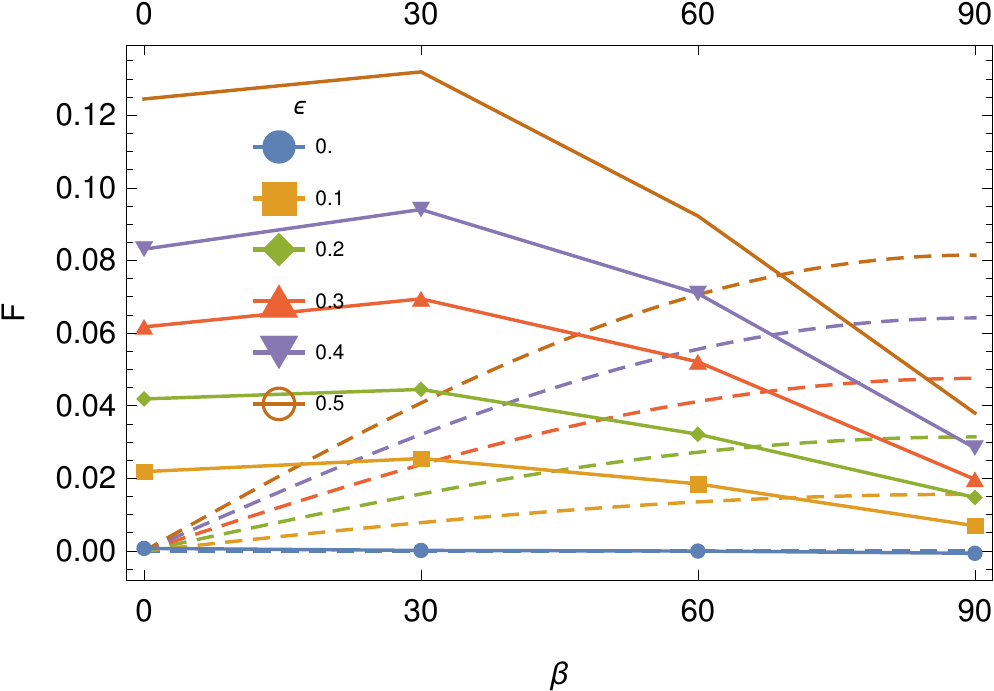}
	\caption{Electromagnetic force induced by a rotating off-centred dipole for different displacements~$\epsilon$ and different angles~$\beta$ for $\alpha=60\degr$, and marked as solid coloured lines. The dashed coloured lines correspond to the vacuum analogue. No fits are shown.}
	\label{fig:force_a60_beta}
\end{figure}

Discrepancies also arise when showing the dependence on the angle~$\beta$, for $\alpha=30\degr$ in Fig.\ref{fig:force_a30_beta} and for $\alpha=60\degr$ in Fig.\ref{fig:force_a60_beta}. The behaviour now deviates significantly from the expectations of a vacuum dipole. The $\sin\beta$ dependence has changed to a more complicated angular dependence we fitted with
expressions containing $\cos\beta, \sin\beta, \cos^2\beta, \sin^2\beta$. As seen from the lots, the fits perform well for $\epsilon \leq 0.3$ but become much less accurate otherwise. For comparison, we also show the vacuum electromagnetic force in dashed coloured lines.

Some fitting formulae are given for the kick depending on the displacement~$\epsilon$ in Table~\ref{tab:fit_force}. 	
For the $\beta$ dependence, we did not found any simple and useful formula so we do not show them.
\begin{table}
\centering
\begin{tabular}{cccc}
\hline
$\alpha$ & $\beta$ & $F_1$ & $F_3$ \\
\hline
\hline
30 &	0  & 0.277 & -0.185 \\
30 &	30 & 0.245 & -0.172 \\
30 &	60 & 0.131 & 0.166 \\
30 &	90 & 0.069  & -0.117 \\
\hline
60 & 	0  & 0.184 & 0.237 \\
60 &	30 & 0.213 & 0.189 \\
60 &	60 & 0.165 & 0.077 \\
60 &	90 & 0.066 & 0.038 \\
\hline
\end{tabular}
\caption{\label{tab:fit_force}Best fit coefficients for the force $F$ expressed as $F_1\,\epsilon + F_3 \,\epsilon^3$ for $\alpha=30\degree$ and $60\degree$.}
\end{table}
We go on by computing the electromagnetic torque from the off-centred force-free dipole.

\section{Electromagnetic torque}
\label{sec:Couple}

Finally, as for the vacuum case, the electromagnetic torque is computed employing the same procedure as in \cite{petri_impact_2019}. This torque is given by integration of the Laplace force on the surface of the star~$\mathcal{S}_*$ such that
\begin{equation}
\label{eq:Couple_Laplace}
\mathbf K = R^3 \, \iint_{\mathcal{S}_*} \mathbf [ \sigma_{\rm s} \, \mathbf n \wedge \mathbf E + ( \mathbf B \cdot \mathbf n ) \, \mathbf i_{\rm s} ] \, d\Omega 
\end{equation}
where $\mathbf n$ is the unit normal to the surface. We identify two contributions to this torque, the first one arising from the surface charge density, $\sigma_{\rm s} = \varepsilon_0 \, [\mathbf{E}] \cdot \mathbf{n}$ and the second from the surface current density $\mu_0 \, \mathbf i_{\rm s} = \mathbf{n} \wedge [\mathbf{B}]$. The square bracket notation $[\mathbf{F}]$ means the jump of the vector field $\mathbf{F}$ across the stellar surface. We also assume that the force-free condition holds inside the star, therefore there is no volume contribution to the torque as by definition $\rho \, \mathbf{E} + \mathbf{j} \wedge \mathbf{B} = \mathbf{0}$ where $(\mathbf{E},\mathbf{B})$ is the electromagnetic field, $\rho$ the charge density and $\mathbf{j}$ the current density inside the star. $\varepsilon_0$ and $\mu_0$ are the vacuum permittivity and permeability. 
The assumption of a force-free neutron star interior means that the electromagnetic force acts only on the stellar surface, admittedly an oversimplified picture. Some non force-free currents could certainly flow inside the star but would require a deeper knowledge of the stellar interior, a task out of the scope of the present study (see \cite{paschalidis_new_2013} for a technique to join a force-free magnetosphere to the MHD interior of the star). Even if the torque could be calculated as an angular momentum flow through a surface surrounding the star, it would require an assumption about the stellar interior because of the need to impose the continuity of the radial magnetic field and the tangential electric field across the surface.

\begin{figure}
	\centering
	\includegraphics[width=0.99\linewidth]{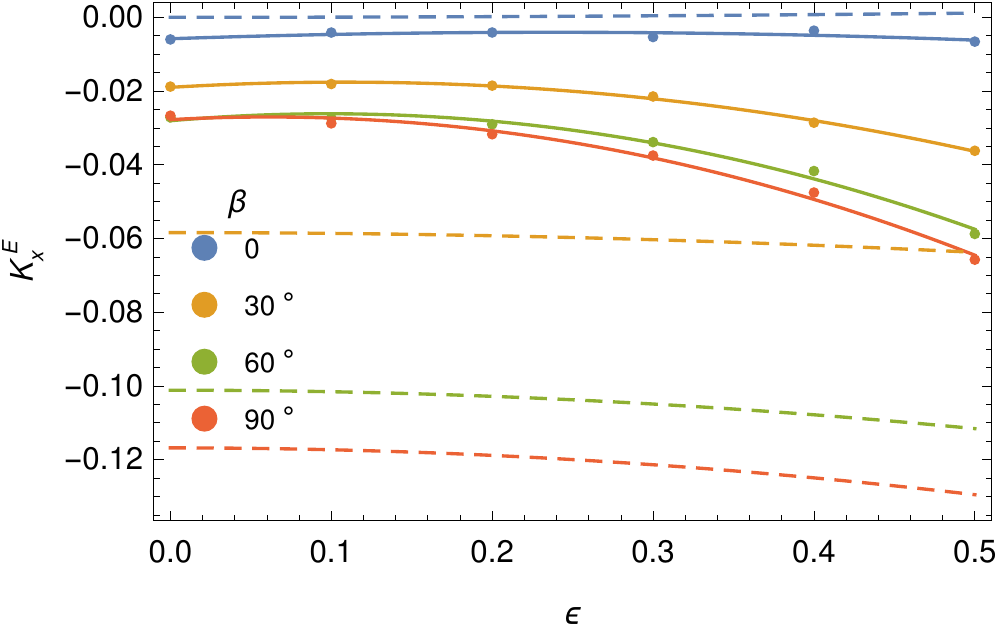}
	\caption{$K_{\rm x}^{\rm E}$ component of the electric torque induced by a rotating off-centred dipole for different displacements~$\epsilon$ and angles~$\beta$ for $\alpha=60\degr$. Coloured dots are from the simulations whereas the solid coloured lines are the fits. The dashed coloured lines correspond to the vacuum analogue divided by a factor~10.}
	\label{fig:couple_E_x_r0.3_a60_beta}
\end{figure}

\begin{figure}
	\centering
	\includegraphics[width=0.99\linewidth]{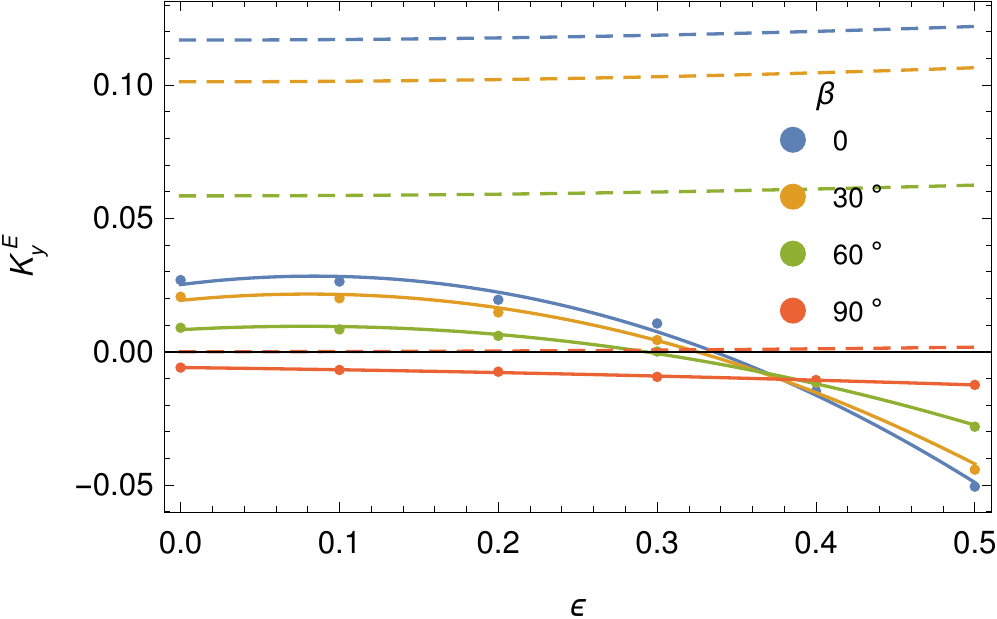}
	\caption{$K_{\rm y}^{\rm E}$ component of the electric torque induced by a rotating off-centred dipole for different displacements~$\epsilon$ and angles~$\beta$ for $\alpha=60\degr$. Coloured dots are from the simulations whereas the solid coloured lines are the fits. The dashed coloured lines correspond to the vacuum analogue divided by a factor~10.}
	\label{fig:couple_E_y_r0.3_a60_beta}
\end{figure}

\begin{figure}
	\centering
	\includegraphics[width=0.99\linewidth]{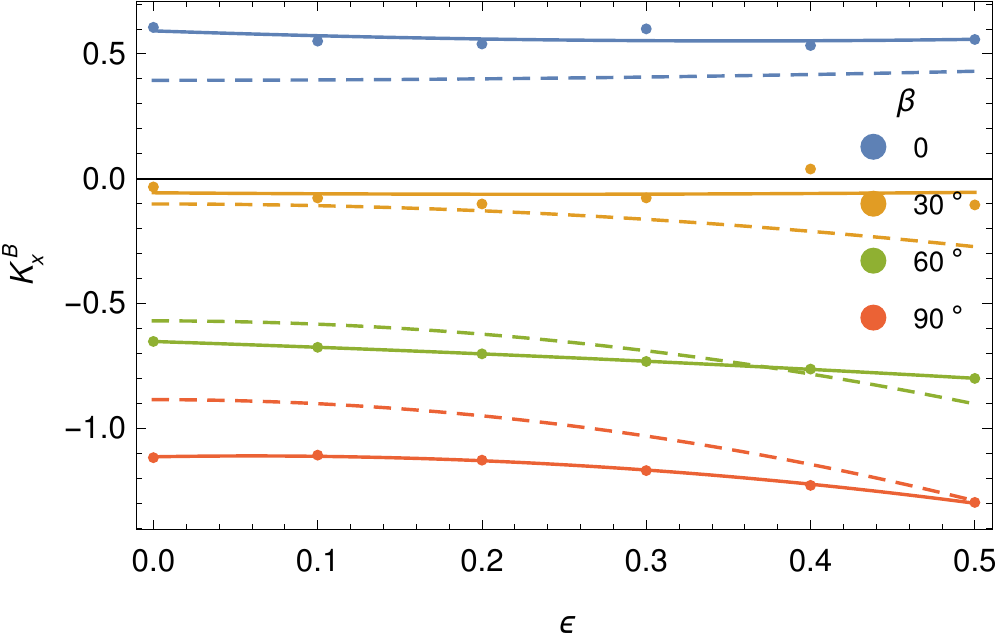}
	\caption{$K_{\rm x}^{\rm B}$ component of the magnetic torque induced by a rotating off-centred dipole for different displacements~$\epsilon$ and angles~$\beta$ for $\alpha=60\degr$. Coloured dots are from the simulations whereas the solid coloured lines are the fits. The dashed coloured lines correspond to the vacuum analogue.}
	\label{fig:couple_B_x_r0.3_a60_beta}
\end{figure}

\begin{figure}
	\centering
	\includegraphics[width=0.99\linewidth]{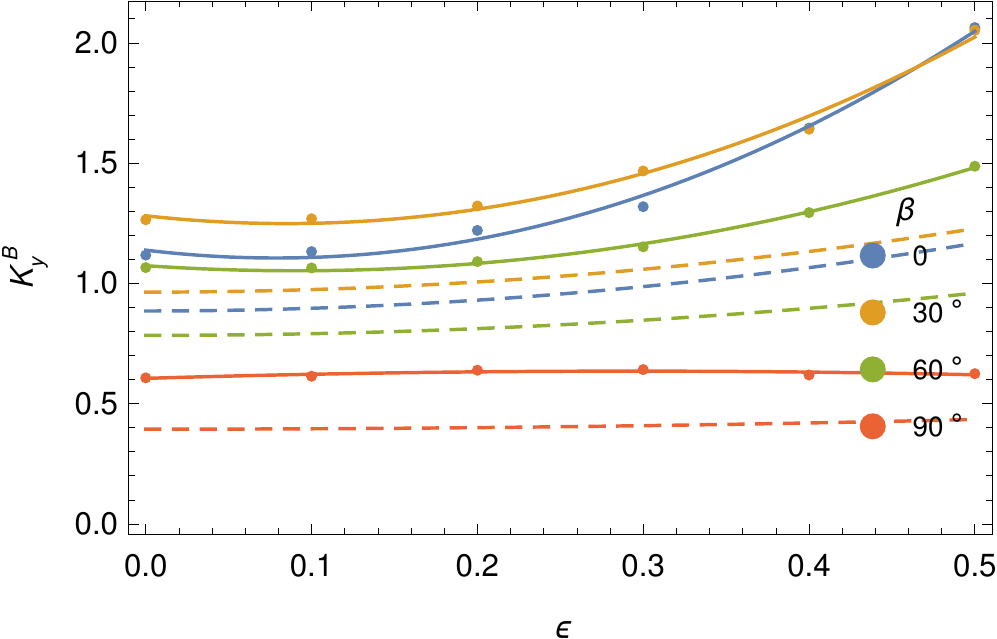}
	\caption{$K_{\rm y}^{\rm B}$ component of the magnetic torque induced by a rotating off-centred dipole for different displacements~$\epsilon$ and angles~$\beta$ for $\alpha=60\degr$. Coloured dots are from the simulations whereas the solid coloured lines are the fits. The dashed coloured lines correspond to the vacuum analogue.}
	\label{fig:couple_B_y_r0.3_a60_beta}
\end{figure}

\begin{figure}
	\centering
	\includegraphics[width=0.99\linewidth]{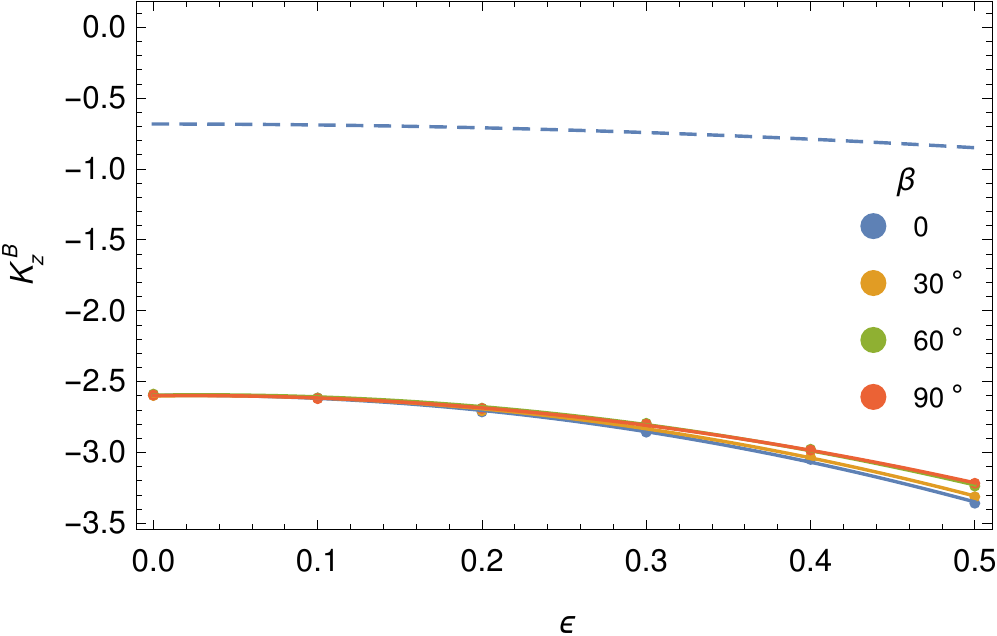}
	\caption{$K_{\rm z}^{\rm B}$ component of the magnetic torque induced by a rotating off-centred dipole for different displacements~$\epsilon$ and angles~$\beta$ for $\alpha=60\degr$. Coloured dots are from the simulations whereas the solid coloured lines are the fits. The dashed coloured lines correspond to the vacuum analogue.}
	\label{fig:couple_B_z_r0.3_a60_depl}
\end{figure}

\begin{figure}
	\centering
	\includegraphics[width=0.99\linewidth]{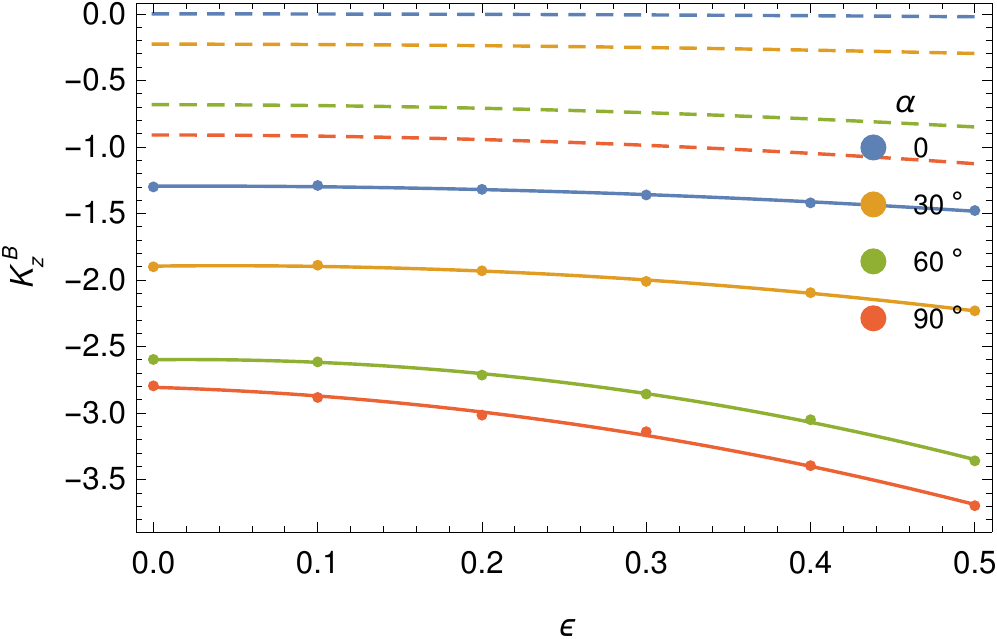}
	\caption{$K_{\rm z}^{\rm B}$ component of the magnetic torque induced by a rotating off-centred dipole for different displacements~$\epsilon$ and angles~$\alpha$. This component is weakly dependent of $\beta$. Coloured dots are from the simulations whereas the solid coloured lines are the fits. The dashed coloured lines correspond to the vacuum analogue.}
	\label{fig:couple_B_z_r0.3_a60_beta}
\end{figure}

We reckon separately the electric and magnetic contributions to this torque, denoting them by $\mathbf{K}^{\rm E}$ and $\mathbf{K}^{\rm B}$. The results are shown individually for the electric torque $K_{\rm x}^{\rm E}$ along the $x$ axis in Fig.~\ref{fig:couple_E_x_r0.3_a60_beta}, the electric torque $K_{\rm y}^{\rm E}$ along the $y$ axis in Fig.~\ref{fig:couple_E_y_r0.3_a60_beta}, the magnetic torque $K_{\rm x}^{\rm B}$ along the $x$ axis in Fig.~\ref{fig:couple_B_x_r0.3_a60_beta}, the magnetic torque $K_{\rm y}^{\rm B}$ along the $y$ axis in Fig.~\ref{fig:couple_B_y_r0.3_a60_beta} and magnetic torque $K_{\rm z}^{\rm B}$ along the $z$ axis in Fig.~\ref{fig:couple_B_z_r0.3_a60_depl}. The latter being relatively insensitive to the $\beta$ angle, we show it $\alpha$ dependence in Fig.~\ref{fig:couple_B_z_r0.3_a60_beta}. Contrary to the kick, we found accurate fits to each of these components with good analytical expressions even for high displacements~$\epsilon\approx0.3$. 

These fits are listed in Table~\ref{tab:fit_torque} for all components of the torque due to the electric part and the magnetic part. The $z$ component of the magnetic torque only weakly depends on $\beta$ as seen from the fits. More relevant are the fits for varying $\alpha$ given in the same Table~\ref{tab:fit_torque} for this component of the torque.
\begin{table}
	\centering
$K_{\rm x}^{\rm E}$
\begin{tabular}{cccc}
\hline
$\beta$ & $T_0$ & $T_1$ & $T_2$ \\
\hline
\hline
0  & -0.006 & 0.015 & -0.031 \\
30 & -0.019 & 0.026 & -0.122 \\
60 & -0.028 & 0.038 & -0.195 \\
90 & -0.028 & 0.024 & -0.195 \\
\hline
\end{tabular} \\
$K_{\rm y}^{\rm E}$
\begin{tabular}{cccc}
\hline
0  & 0.025 & 0.075 & -0.447 \\
30 & 0.019 & 0.059 & -0.362 \\
60 & 0.008 & 0.033 & -0.209 \\
90 & -0.006 & -0.007 & -0.012 \\
\hline
\end{tabular} \\
$K_{\rm x}^{\rm B}$
\begin{tabular}{cccc}
\hline
0  & 0.592 & -0.225 & 0.316 \\
30 & -0.057 & -0.051 & 0.109 \\
60 & -0.652 & -0.217 & -0.156 \\
90 & -1.114 & 0.117 & -0.979 \\
\hline
\end{tabular} \\
$K_{\rm y}^{\rm B}$
\begin{tabular}{cccc}
\hline
0  & 1.138 & -0.833 & 5.307 \\
30 & 1.281 & -0.763 & 4.499 \\
60 & 1.074 & -0.465 & 2.564 \\
90 & 0.605 & 0.206 & -0.352 \\
\hline
\end{tabular} \\
$K_{\rm z}^{\rm B}$
\begin{tabular}{cccc}
\hline
$\alpha$ & 1 & $\epsilon$ & $\epsilon^2$ \\
\hline
\hline
0  & -1.294 & 0.036 & -0.827 \\
30 & -1.896 & 0.142 & -1.63 \\
60 & -2.599 & 0.129 & -3.26 \\
90 & -2.807 & -0.371 & -2.781 \\
\hline
\end{tabular}
\caption{\label{tab:fit_torque}Best fit coefficients for the torque components $\mathbf{K}$ expressed as $T_0 + T_1\,\epsilon + T_2 \,\epsilon^2$ for $\alpha=60\degree$ except for $K_{\rm z}^{\rm B}$ which is almost independent of $\beta$ so we show it for varying $\alpha$.}
\end{table}
As expected from the spin-down rate, the $z$-component of the FFE torque is stronger than for the vacuum rotator and exists also for an aligned rotator with $\alpha=0\degree$.

We conclude this paper by a last section about possible consequences for neutron stars in binary systems.

\section{Impact on binary and isolated neutron stars}
\label{sec:Binaire}

An off-centred rotating force-free magnetic dipole can have some impact on the orbital evolution of a neutron star binary but also on isolated neutron stars. In this section, we re-explore such questions in the light of our new results compared to the vacuum case presented in \cite{petri_impact_2019}.

\subsection{Binary neutron star eccentricity}

The spin-down luminosities and the electromagnetic kicks induced by an off-centred force-free dipole are comparable in magnitude to the one obtained in vacuum by \cite{petri_impact_2019}. Its impact for binary neutron stars orbit eccentricities has been discussed in depth by \cite{petri_impact_2019} for a vacuum rotator. The contribution of a force-free plasma as the one shown in the present paper would give similar results. However, the geometrical dependence of the spin-down, electromagnetic force and torque are drastically affected by this plasma. The variation with respect to $\alpha$ and $\beta$ have nothing comparable to the vacuum case. This has profound implications for the evolution of the spinning neutron star geometry, that is the evolution of its obliquity~$\alpha$ and therefore also on the braking index. The line of sight will evolve accordingly with a secular change in the multi-wavelength pulse profiles. Here however we only focus on the binary orbital evolution.

Let us summarize the binary neutron star orbital evolution for force-free magnetospheres. Neutron star binaries are expected to relax to almost circular orbits with very low eccentricities~$e\approx 0$ due to mass transfer and tidal circularisation. Nevertheless, large eccentricities $e\gtrsim0.3$ can be produced by supernova explosions when a substantial fraction of the binary mass is lost. The electromagnetic kick produced by an off-centred force-free dipole is able to modify the orbital eccentricity, sometimes generating moderate to large eccentricities in neutron star binaries as we demonstrate below.
	
When both neutron stars of the binary, located respectively at a vector position $\mathbf{r}_1$ and $\mathbf{r}_2$, are subject to kicks $\mathbf{F}_1$ and $\mathbf{F}_2$ (the origin of which is not necessarily electromagnetic), the associated two body problem reduces to the Stark problem also called the accelerated Kepler problem \citep{namouni_accelerated_2007}. The derivation is shown by \cite{petri_impact_2019}. Using the equivalence with a one body problem as in the case of no kicks, the binary neutron star relative separation $\mathbf{r} = \mathbf{r}_2 - \mathbf{r}_1$ satisfies
\begin{equation}
\label{eq:AcceleratedKeplerProblem}
\frac{d^2\mathbf{r}}{dt^2} = - G \, (m_1+m_2) \, \frac{\mathbf{r}}{r^3} + \mathbf{A}
\end{equation}
$m_1$ being the mass of the star located at $\mathbf{r}_1$ and $m_2$ the mass of its companion located at $\mathbf{r}_2$.
The additional acceleration $\mathbf{A}$ is expressed by
\begin{equation}
\label{eq:Acceleration}
\mathbf{A} = \frac{\mathbf{F}_2}{m_2} - \frac{\mathbf{F}_1}{m_1} .
\end{equation}
It is straightforward to check that this acceleration is constant in direction and time for an electromagnetic kick produced by the rotation of an off-centred dipole. Therefore \eqref{eq:AcceleratedKeplerProblem} is fully integrable according to \cite{lantoine_complete_2011}. Focusing on bound orbits for a constant acceleration~$\mathbf{A}$, solutions for eccentricity excitation are given in \cite{namouni_origin_2005, namouni_accelerated_2007}.
	
When the acceleration~$\mathbf{A}$ acts constantly in time, an orbit starting from zero eccentricity $e(t=0)=0$ varies periodically in time according to 
\begin{equation}
\label{eq:EvolutionEccentricite}
e(t) = \left| \sin i_0 \, \sin\left( \frac{3\,A}{2\,\Omega\,a} \, t\right) \right|
\end{equation}
where $a$ is the semi-major axis, $A=\mathbf{\|A\|}$ is the acceleration produced by the neutron stars themselves due to the electromagnetic kick~$F_i$, $i_0$ is the inclination angle between the acceleration vector~$\mathbf{A}$ and the orbital angular momentum vector and $\Omega = \sqrt{G \, (m_1+m_2)/a^3}$ is the keplerian frequency. The typical timescale of eccentricity excitation is then
\begin{equation}
\label{eq:TempsExcentricite}
T_e = \frac{\upi\,\Omega\,a}{3\,A} =  \frac{\upi}{3\,A} \, \sqrt{\frac{G \, (m_1+m_2)}{a}}.
\end{equation}
Because the star is spinning down, the acceleration does not remains constant in time, it decreases significantly after the characteristic age time scale~$\tau_{\rm c}$. The estimate for a constant in time acceleration must be revised taking into account $\tau_{\rm c}$. A good guess for the actual binary eccentricity at the end of the acceleration process is therefore
\begin{equation}
\label{eq:EccentriciteFinale}
e(\tau_{\rm c}) = \left| \sin i_0 \, \sin\left( \frac{3\,A \, \tau_{\rm c}}{2\,\Omega\,a} \right) \right| .
\end{equation}
The upper limit for the eccentricity is achieved after a full excitation period~$T_e$ and according to eq.~\eqref{eq:EvolutionEccentricite} amounts to $\sin i_0$. This time~$T_e$ has to be compared to other typical time scales like the true age of the binary and the electromagnetic spin-down time scale. The eccentricity depends only on the neutron star spinning period $P$ and on the orbital period $P_{\rm orb}$ but not on the period derivative $\dot{P}$. The typical eccentricity therefore becomes
\begin{equation}
e(\tau_{\rm c}) = \left| \sin i_0 \, \sin\left( \frac{18\,\upi^{5/3}}{5\times 2^{1/3}} \, \epsilon \, \frac{I\,P^{-2} \, P_{\rm orb}^{1/3}}{m_1\,c\,\sqrt{G\,(m_1+m_2)}} \right) \right| .
\end{equation}
For low eccentricities, it reduces to
\begin{subequations}
\begin{align}
e(\tau_{\rm c}) & \approx \frac{18\,\upi^{5/3}}{5\times 2^{1/3}} \, \epsilon  \, | \sin i_0 | \, \frac{I\,P^{-2} \, P_{\rm orb}^{1/3}}{m_1\,c\,\sqrt{G\,(m_1+m_2)}} \\
	& \approx 1.5 \times 10^{-5} \, \epsilon \, | \sin i_0 \, | \left( \frac{P}{1~\textrm{s}}\right)^{-2} \, \left(\frac{P_{\rm orb}}{1~\textrm{day}}\right)^{1/3}
	\end{align}
\end{subequations}
revealing a simple scaling with $P$ and $P_{\rm orb}$ as
\begin{equation}
\label{eq:eccentricity}
e \propto \epsilon \, P^{-2} \, P_{\rm orb}^{1/3} .
\end{equation}
Compared to the vacuum case, the formal dependence is similar expect for the geometrical factor depending on the angles~$\alpha,\beta,\delta$, factor now deduced from the simulation results.

\subsection{Isolated neutron star kick}

According to several observations investigations, pulsar proper motion seems to be almost aligned with their rotation axis \citep{hobbs_statistical_2005} \cite{johnston_evidence_2005}. The electromagnetic kick has been suggested as an alternative scenario to mechanisms imprinted a large kick velocity right a the birth of the neutron star. See for instance \cite{lai_pulsar_2001} for a discussion and a more recent extension by \cite{wang_neutron_2006}. \cite{kojima_kick_2011} studied a particular case of magnetic dipole + quadrupole electromagnetic kick showing the evolution of the kick velocity with time. The final kick velocity depends on the relative magnitude of te quadrupole versus dipole and on their respective orientation rather than on the magnitude itself.

\begin{figure}
	\centering
	\includegraphics[width=\linewidth]{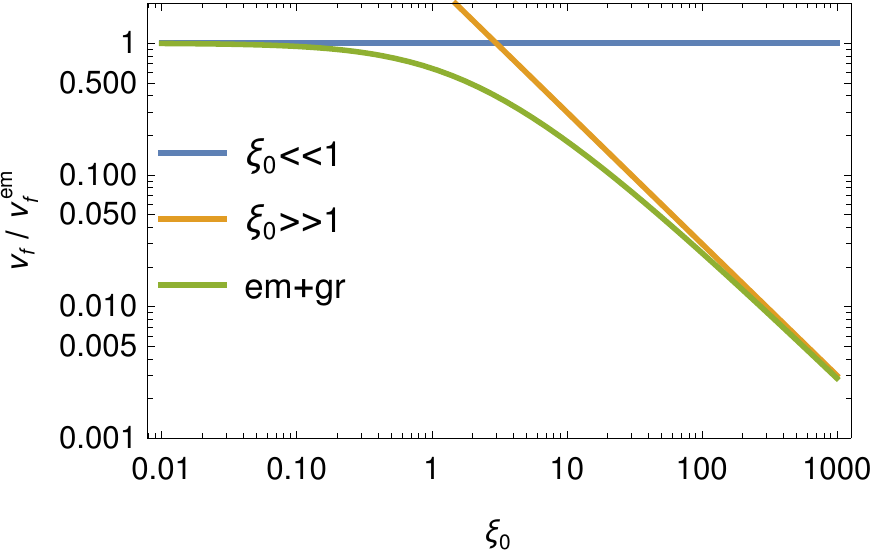}
	\caption{Final kick velocity comparing the pure electromagnetic case $v_f^{\rm em}$ to the gravitational+electromagnetic case $v_f$ depending on the ratio $\xi_0 = L_{\rm gw}/L_{\rm em}$.}
	\label{fig:kickfinal}
\end{figure}

From the electromagnetic kick expressions found in the previous section, an efficient recoil requires a fast spinning neutron star possessing a large off-centred dipole. Large magnetic field strengths are not required but their decrease the timescale of the kick. The spin-down rate is mostly accounted for by the magnetodipole radiation losses. However, in the early youth of the star, gravitational radiation can be substantial. The net effect on the kick velocity decreases significantly in such cases as shown by \cite{lai_pulsar_2001}. We re-examine this situation in the following lines. As in the vacuum case, the force can be estimated by $F(t)\approx a \, \epsilon \, L_{\rm em}(t)/c$ disregarding geometric factors involving the angles $\alpha,\beta,\delta$. Assuming a star starting at rest at birth, the kick velocity after a time $t$ is
\begin{equation}\label{eq:vitesse}
v(t) = \int_{0}^{t} \frac{F(t)}{M} \, dt = \frac{\epsilon}{M\,c} \, \int_{0}^{t} a\,L_{\rm em}(t) \, dt .
\end{equation}
The argument put forward in this calculation assumes implicitly that the rotation axis is fixed with respect to the magnetic field configuration. However, electromagnetic radiation produces also a torque responsible for the alignment between rotation and magnetic dipole axis. It has been shown in \cite{petri_magnetic_2020} that the timescale for this alignment is of the same order of magnitude as the electromagnetic quick time scale, therefore, to a good approximation, we can neglect this alignment for the estimate of the final kick velocity. Even if the alignment is properly taken into account, the kick obtained remains almost the same, see Eq.(32b) in \cite{petri_magnetic_2020}.
If the spin-down is fully of electromagnetic origin, then $L_{\rm em}(t) = -I\,\Omega\,\dot{\Omega}$ where $I$ is the stellar moment of inertia and the kick velocity becomes for an initial spin of $\Omega_0$
\begin{equation}\label{eq:v_t}
v(t) = \frac{I\,\epsilon\,R}{3\,M\,c^2} \, (\Omega_0^3 - \Omega(t)^3) .
\end{equation}
Taking a moment of inertia for a homogeneous sphere, $I=\frac{2}{5}\,M\,R^2$ we get
\begin{equation}\label{eq:v_t2}
v(t) = \frac{2\,c}{15} \, \epsilon \, (a_0^3 - a(t)^3)
\end{equation}
with $a_0 = \Omega_0\,R/c$. The final velocity at large times is 
\begin{subequations}\label{eq:vfinale}
\begin{align}
v_{\rm f}^{\rm em} & = \frac{2\,c}{15} \, \epsilon \, a_0^3 \\
 & = \numprint{64}~\SIunits{\kilo\meter\per\second} \, \left(\frac{\epsilon}{0.1} \right) \, \left( \frac{R}{\numprint{12}~\SIunits{\kilo\meter}}\right)^3 \, \left( \frac{P}{1~\SIunits{\milli\second}}\right)^{-3} .
\end{align}
\end{subequations}
We observe that the final velocity is independent of the magnetic field strength, it only depends on the initial rotation period of the pulsar at birth. High kick velocities can therefore only be explained by very high initial rotation rates, in the sub-millisecond range. The situation gets worth if gravitational radiation is taking into account. Indeed, the rotational history of the star then follows
\begin{equation}\label{eq:Omegadot}
\dot{\Omega} = - (k_1 \, \Omega^3 + k_2 \, \Omega^5) .
\end{equation}
The constant coefficients for magneto-dipole losses and gravitational wave are respectively
\begin{subequations}
\begin{align}
k_1 \, I & = \frac{8\,\upi}{3\,\mu_0\,c^3} \, B^2 \, R^6 \\
k_2 \, I & = \frac{32}{5} \, \frac{G}{c^5} \, \mu^2 \, I_{zz}^2
\end{align}
\end{subequations}
where we used the gravitational luminosity for a deformed neutron star with ellipticity $\mu$ \citep{shapiro_black_1983}. The final kick velocity then becomes with only the electromagnetic spin-down contributing to the kick $L_{\rm em} = k_1\,I\,\Omega^4$ and using Eq.~\eqref{eq:vitesse}
\begin{equation}\label{eq:v_f}
v_{\rm f} = \frac{\epsilon}{M\,c^2} \, \int_{0}^{t} \, \frac{\Omega\,R}{c} \, k_1\,I\,\Omega^4 \, dt = -\frac{\epsilon\,I\,R}{M\,c^3} \, \int_{\Omega_0}^{0} \frac{k_1\,\Omega^5}{k_1\,\Omega^3+k_2\,\Omega^5} \, d\Omega .
\end{equation}
Using the typical value for the homogeneous sphere moment of inertia, we arrive at an expression similar to \cite{kojima_kick_2011} but with different notations
\begin{equation}\label{eq:vfinalgravitation}
v_{\rm f} = \frac{2\,c}{5} \, \epsilon \, \frac{a_0^3}{\xi_0} \, \left[ 1 - \frac{\arctan \sqrt{\xi_0}}{\sqrt{\xi_0}} \right] .
\end{equation}
The parameter $\xi_0 = k_2\,\Omega_0^2/k_1 = L_{\rm gw}/L_{\rm em}$ controls the initial ratio between gravitational~$L_{\rm gw}$ and electromagnetic~$L_{\rm em}$ luminosity. 
The final kick is compared to the pure electromagnetic kick $v_f^{\rm em}$ in Fig.~\ref{fig:kickfinal}.
If gravitational radiation is negligible, $\xi_0\ll1$ and we retrieve expression~(\ref{eq:vfinale}). In the opposite case where gravitational radiation is dominant, $\xi_0\gg1$ the asymptotic kick velocity tends to zero according to
\begin{equation}\label{eq:vfinalgravitation2}
\frac{v_f}{c} = \frac{2}{5} \, \epsilon \, \frac{a_0^3}{\xi_0} .
\end{equation}
If the gravitational wave spin-down dominates the electromagnetic losses at birth, the final kick velocity will be drastically reduced by a factor $\xi_0$. This is because gravitational wave emission does not produce any recoil. In order to get a large recoil, the electromagnetic force must act on a long time scale, similar to the Poynting spin-down time scale that cannot be achieved if gravitational wave emission is preponderant. Favourable kick velocities are obtained for $\xi_0\ll1$ which gives a constrain on ellipticity~$\mu$ versus magnetic dipole field~$B$ assuming an initial period~$P_0$. The characteristic age for a quadrupole is also $\xi_0$ smaller than the characteristic age for a dipole.

In principle, gravitation radiation can also produce a recoil of the star. For instance, this mechanism imprints a substantial kick in binary mergers of compact objects. Gravitational recoil requires at least some next to leading order quadrupole and octupole moments interactions (mass and/or current) to transfer linear momentum to the star \citep{thorne_multipole_1980, maggiore_gravitational_2007}. For black holes, \cite{bekenstein_gravitational-radiation_1973} estimated the recoil velocity order of magnitude around typically 300~km/s. For binary black hole mergers, the dynamics is different but the recoil remains of the same order of magnitude, although slightly higher, about 1000~km/s \citep{fitchett_influence_1983}. 
For an axisymmetric system \cite{bekenstein_gravitational-radiation_1973} found a simple expression for the recoil. This however cannot be applied to neutron stars because gravitational wave emission requires breaking of this axisymmetric configuration. To lowest order, the quadrupole mass moment~$Q_{ik}$, the quadrupole current moment~$S_{ik}$ and the octupole mass moment~$Q_{ijk}$ produce together a recoil given for instance in \cite{blanchet_analyzing_2019} by
\begin{equation}\label{eq:recoil}
 \frac{dP_i}{dt} = \frac{G}{c^7} \, \left( \frac{2}{63} \, \frac{d^3 Q_{ij}}{dt^3} \, \frac{d^4 Q_{ijk}}{dt^4} + \frac{16}{45} \, \varepsilon_{ijk} \, \frac{d^3 Q_{j\ell}}{dt^3} \, \frac{d^3 S_{k\ell}}{dt^3} \right) .
\end{equation}
We use this expression not to solve for the exact problem but to give some orders of magnitude of the expected recoil from an isolated neutron star radiating gravitational waves. In orders of magnitude, the terms in bracket scale as $F_{\rm gr} \propto M^2\,R^5\,\Omega^7$ whereas the gravitation luminosity scales as $L_{\rm gr} \propto M^2\,R^4\,\Omega^6$. We therefore obtain a very similar expression to the dipole-quadrupole magnetic field, namely that
\begin{equation}\label{eq:force_gr}
F_{\rm gr} = \kappa \, \frac{a\,L_{\rm gr}}{c}
\end{equation}
where $\kappa$ encompasses the magnitude of the mass octupole and current quadrupole terms. The final gravitational kick velocity would resemble the final electromagnetic kick velocity in Eq.~\eqref{eq:vfinale} by replacing $\epsilon$ with $\kappa$. However, this next to leading order contribution from $Q_{ijk}$ and $S_{ij}$ is very small $\kappa\ll1$ because the departure from spherical symmetry is weak. Consequently, the kick imprinted by gravitational radiation alone remains negligible, a factor $\kappa/\epsilon$ smaller than for the electromagnetic kick above mentioned. To find and estimation of the neutron star deformation, let us assume that its distortion is due to its own magnetic field \citep{bonazzola_gravitational_1997}. The ellipticity is therefore, introducing the magnetic distortion factor~$\beta$, synthesizing the magnetic stress on the star shape, and taking the ratio between magnetic energy and gravitational potential energy
\begin{equation}\label{eq:ellipticite}
 e = \beta \, \frac{4\,\upi}{3\,\mu_0} \, \frac{B^2\,R^4}{G\,M^2} = \numprint{e-12} \, \beta \, \left( \frac{B}{\numprint{e8}~\SIunits{\tesla}} \right)^2 \ll 1 .
\end{equation}
Even if the magnetic distortion factor~$\beta$ can be as large as $1000$, as found by \cite{bonazzola_gravitational_1997}, it is insufficient to significantly distort the star.

An absolute upper limit for the kick velocity is obtained at the mass shedding limit $\Omega_{\rm k} = \sqrt{G\,M/R^3}$, assuming Newtonian gravity. General-relativistic corrections discussed by \cite{friedman_implications_1989,haensel_equation_1995} decrease this value by approximately~$2/3$. This corresponds then to a spin parameter
\begin{subequations}\label{eq:a_k}
\begin{align}
a_k & = \frac{2}{3} \, \frac{\Omega_k\,R}{c} = \sqrt{\frac{G\,M}{R\,c^2}} \\
 & = 0.278 \, \left( \frac{M}{1.4~M_\odot} \right)^{1/2} \, \left( \frac{R}{\numprint{12}~\SIunits{\kilo\meter}}\right)^{-1/2} .
\end{align}
\end{subequations}
We have
\begin{subequations}\label{eq:vfinalmax}
\begin{align}
v_f^{\rm max} & = \frac{2}{15} \, \epsilon \, \left( \frac{G\,M}{R\,c^2}\right)^{3/2} \\
 & = \numprint{86}~\SIunits{\kilo\meter\per \second} \, \left(\frac{\epsilon}{0.1}\right) \, \left( \frac{M}{1.4~M_\odot} \right)^{3/2} \, \left( \frac{R}{\numprint{12}~\SIunits{\kilo\meter}}\right)^{-3/2} .
\end{align}
\end{subequations}
A high initial kick velocity requires a large off-centred dipole with $\epsilon\lesssim1$. More generally speaking, it means that multipolar components must be as large as or even larger than the dipolar component. We could imagine a less restrictive geometry by relaxing the off-centred dipole and choose a dipole+quadrupole configuration leaving the relative magnetic strength between dipole and quadrupole as a free parameter. The electromagnetic kick scenario remains however interesting because it naturally explained the spin-proper motion alignment observed in many neutron stars \citep{johnston_evidence_2005}.

The importance of gravitational wave emission in the early phases depend on the ellipticity of the neutron star which is unfortunately ill constrained with upper limits for isolated radio pulsars given by $\mu \approx \numprint{e-4}-\numprint{e-6}$ \citep{aasi_gravitational_2014}. Recent searches for continuous gravitational waves seem even to constrain typical values to be less than around $\mu \approx 10^{-8}$ \citep{abbott_searches_2019} even for millisecond pulsars \citep{abbott_gravitational-wave_2020}. Comparing to magneto-dipole losses, we get
\begin{subequations}\label{eq:LgwsLem}
\begin{align}
\frac{L_{\rm gw}}{L_{\rm em}} & = \frac{k_2\,\Omega_0^2}{k_1} = \frac{48}{125} \, \frac{\mu_0}{c^2} \, \frac{G\,M^2\,\Omega^2\,\mu^2}{B^2\,R^2} \\
& = \numprint{7.6} \, \left( \frac{P}{1~\SIunits{\milli\second}}\right)^{-2} \, \left( \frac{R}{12~\SIunits{\kilo\meter}}\right)^{-2} \, \left( \frac{B}{10^8 \, \SIunits{\tesla}} \right)^{-2} \, \left( \frac{\mu}{\numprint{e-5}}\right)^2 . \nonumber
\end{align}
\end{subequations}
For isolated radio pulsars with typical magnetic field strength of $10^8 \, \SIunits{\tesla}$ and initial period of $P_0=1$~\SIunits{\milli\second}, an ellipticity stronger than \numprint{e-5} generates a large gravitational spin-down luminosity, dominating the electromagnetic spin-down in the early phase of the neutron star. 

In any case, the off-centred dipole cannot explain the fastest moving neutron stars even for large displacements up to almost the surface if gravitational wave emission is significant in the early stage as expected from neutron star formation scenario. This is due to the constrain on $\epsilon \lesssim1$ that imposes a maximum value for the quadrupole. If this condition is released, for instance for a quadrupolar component not related to the dipole \citep{kojima_kick_2011}, we would expect higher kick velocities depending on the relative strength between magnetic dipole and quadrupole. We explore this issue in another work.

\section{Conclusions}
\label{sec:Conclusion}

We performed accurate time-dependent numerical simulations of off-centred force-free rotating dipoles scanning a full range of geometrical parameters. We found that the off-centring slightly increases the spin-down luminosity compared to a centred dipole. We fitted this enhancement by a simple expression quadratic in the displacement $\epsilon$. The associated electromagnetic kick and torques have been computed. Our new results show that the magnetospheric plasma has but only little effect compared to the vacuum case. The main difference arises in the formal dependence on the geometry but qualitatively the conclusions presented in \cite{petri_radiation_2016} remain valid. The impact on magnetic field line structures, spin-down luminosities, induced electromagnetic forces and torques have been outlined. The geometrical dependence on the dipole orientation is more involved than for the vacuum case. All angles modify the luminosity, the force and the torque. The electromagnetic kick could have an impact on the orbital evolution of binary neutron stars as was already the case for the vacuum off-centred dipole. For force-free off-centred dipole, we expect similar behaviours as for vacuum dipoles. We also re-explored the question about the velocity kick of isolated neutron stars and show that it cannot easily explain the highest proper motion because of the constrain on the magnetic moment displacement.

One interesting possibility to extend this work releases the assumption of an off-centred dipole, replacing it by a dipole+quadrupole configuration. This alleviates the limit on the quadrupole component with respect to the dipole, increasing the maximum reachable velocity kick. This idea will be detailed in another work.

\section*{Acknowledgements}

This work is also supported by the CEFIPRA grant IFC/F5904-B/2018.  We would like to acknowledge the High Performance Computing center of the University of Strasbourg for supporting this work by providing scientific support and access to computing resources. Part of the computing resources were funded by the Equipex Equip@Meso project (Programme Investissements d'Avenir) and the CPER Alsacalcul/Big Data.

\section*{Data availability}

The data underlying this article will be shared on reasonable request to the corresponding author.










\bsp	
\label{lastpage}
\end{document}